\documentclass[conference]{IEEEtran}
\usepackage{cite}
\IEEEoverridecommandlockouts
\usepackage{cite}
\usepackage{amsmath,amssymb,amsfonts}
\usepackage{algorithmic}
\usepackage{graphicx}
\usepackage{textcomp}
\usepackage{xcolor}
\usepackage{hyperref}
\usepackage{makecell}
\usepackage{multirow}
\usepackage{algorithm}
\usepackage{booktabs}
\usepackage{algorithmic}
\usepackage{bm}
\usepackage{amsmath,amssymb,amsthm}

\usepackage{balance}

\hypersetup{
    colorlinks=true, % Use colored links instead of boxes
    linkcolor=black,  % Color of internal links
    citecolor=black, % Color of citation links
    filecolor=black, % Color of file links
    urlcolor=black    % Color of external links
}

\def\BibTeX{{\rm B\kern-.05em{\sc i\kern-.025em b}\kern-.08em
    T\kern-.1667em\lower.7ex\hbox{E}\kern-.125emX}}
\begin{document}

% \title{Robust and Attack-Resilient Decentralized Quantum Kernel Learning for Noisy Environments}

% \title{The Robustness of Decentralized Quantum Kernel Learning for Noisy and Adversarial Environments}

\title{RDQKL: Robust Decentralized Quantum Kernel Learning for Noisy and Adversarial Environment}

\author{
\IEEEauthorblockN{
    Wenxuan Ma\IEEEauthorrefmark{2}\IEEEauthorrefmark{3},
    Kuan-Cheng Chen\IEEEauthorrefmark{4}\IEEEauthorrefmark{5}\IEEEauthorrefmark{1},
    Shang Yu\IEEEauthorrefmark{6}\IEEEauthorrefmark{5},
    Mengxiang Liu\IEEEauthorrefmark{4},
    and Ruilong Deng\IEEEauthorrefmark{2}\IEEEauthorrefmark{3}\IEEEauthorrefmark{1}
}
\IEEEauthorblockA{\IEEEauthorrefmark{2}College of Control Science and Engineering, Zhejiang University, Hangzhou, China}
\IEEEauthorblockA{\IEEEauthorrefmark{3}State Key Laboratory of Industrial Control Technology, Zhejiang University, Hangzhou, China}
\IEEEauthorblockA{\IEEEauthorrefmark{4}Department of Electrical and Electronic Engineering, Imperial College London, London, UK}
\IEEEauthorblockA{\IEEEauthorrefmark{5}Centre for Quantum Engineering, Science and Technology (QuEST), Imperial College London, London, UK}
\IEEEauthorblockA{\IEEEauthorrefmark{6}Blackett Laboratory, Department of Physics, Imperial College London, London, UK}

\thanks{* Corresponding Authors: kuan-cheng.chen17@imperial.ac.uk and dengruilong@zju.edu.cn}
}

\maketitle

\begin{abstract}
This paper proposes a general decentralized framework for quantum kernel learning (QKL). It has robustness against quantum noise and can also be designed to defend adversarial information attacks forming a robust approach named RDQKL. We analyze the impact of noise on QKL and study the robustness of decentralized QKL to the noise. By integrating robust decentralized optimization techniques, our method is able to mitigate the impact of malicious data injections across multiple nodes. Experimental results demonstrate that our approach maintains high accuracy under noisy quantum operations and effectively counter adversarial modifications, offering a promising pathway towards the future practical, scalable and secure quantum machine learning (QML).
% in the near future. 
\end{abstract}


\section{Introduction}
\label{sec:intro}
Quantum computing harnesses the principles of superposition and entanglement to enable computational paradigms that may surpass classical methods in certain complex tasks, including combinatorial optimization\cite{khairy2020learning, zhou2020quantum, chen2025resource, chen2024noise}, quantum chemistry simulation\cite{kokail2019self, tilly2022variational, chen2024quantum}, machine learning\cite{abbas2021power, yu2024shedding}, sustainability\cite{mohamed2025quantum, ho2024quantum, lin2024quantum}, and drug discovery\cite{yu2023universal, cao2018potential}. These foundational quantum phenomena have spurred the development of quantum algorithms and protocols with the potential to transform scientific and technological fields at scale. One particularly promising application is quantum kernel learning (QKL) \cite{biamonte2017quantum}, in which classical data are embedded into high-dimensional Hilbert spaces via quantum feature maps, allowing similarities (or kernels) to be computed through measured overlaps of quantum states\cite{havlivcek2019supervised,chen2024validating,chen2024quantum3}. Under certain assumptions, these quantum embeddings can access exponentially large feature spaces, potentially enabling more expressive models than purely classical approaches\cite{huang2021power}.

\begin{figure}[!t]
\begin{center}
\centerline{\includegraphics[width=\columnwidth]{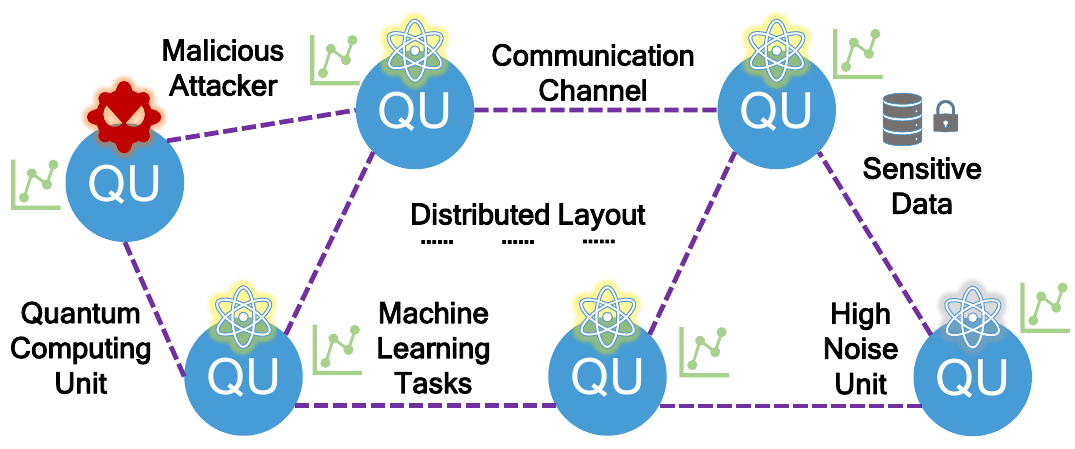}}
% \vskip -0.1in
\caption{A schematic overview of the application scenario. Multiple quantum computing units (QU) collaborate on a shared machine learning task to accelerate the training or preserve the data privacy. In this open environment, certain nodes may suffer from high noise or malicious behavior, potentially sabotaging the training process.}
\label{fig_s}
\end{center}
\end{figure}

Despite these prospects, the performance of quantum kernel methods on near-term devices can be severely hindered by noise processes such as depolarizing errors\cite{hubregtsen2022training}, which degrade quantum states and thereby undermine the expected performance gains\cite{thanasilp2024exponential, wang2025power}. Simultaneously, modern computational infrastructures—including quantum hardware\footnote{https://www.newscientist.com/article/2457325-ibm-will-release-the-largest-ever-quantum-computer-in-2025/}—are becoming increasingly decentralized to address bandwidth constraints, data-privacy regulations, and collaborative research needs\cite{mcmahan2017communication, nedic2009distributed}. This paradigm shift gives rise to decentralized (or federated) quantum learning environments\cite{dean2012large, zhang2021survey}, where multiple quantum computing units collaborate on a shared task for training acceleration or maintaining local datasets just as shown in Fig.\ref{fig_s}.

% In quantum kernel learning, classical data are mapped into high-dimensional Hilbert spaces through carefully designed quantum feature maps, and similarities among data points are computed by measuring overlaps between corresponding quantum states \cite{havlivcek2019supervised}. In principle, this embedding can represent data in exponentially large spaces, enabling more expressive models than those achievable in purely classical settings \cite{huang2021power}. However, on currently available quantum devices, the expected advantages of quantum kernel methods are frequently undermined by pervasive noise sources such as depolarizing errors \cite{hubregtsen2022training}. These noise processes degrade the distinguishability of quantum features and ultimately limit the performance gains that quantum kernels can deliver in practice ~\cite{thanasilp2024exponential, wang2025power}.

However, deploying QKL in decentralized settings introduces unique reliability and security challenges. First, quantum noise typically varies across different hardware nodes; poorly shielded circuits at certain locations may exhibit higher error rates, skewing kernel evaluations and impairing model training. Second, adversarial or faulty nodes can deliberately—or inadvertently—inject corrupted information, further undermining the learning process. These dual challenges of noise and malicious interference demand new methodological safeguards to assure the efficiency and reliability of decentralized QKL.
% is to be both efficient and reliable.

Motivated by these gaps, we analyze the impact of noise on QKL and investigate the robustness of decentralized QKL to the noise. By integrating robust decentralized optimization technique, we propose a Robust Decentralized Quantum Kernel Learning (RDQKL) framework that jointly addresses noise degradation and adversarial threats in a unified manner. Specifically, the decentralized protocol ensures that nodes with lower noise levels guide updates for higher-noise nodes without sacrificing the overall convergence. Through robust aggregation rules with clipping mechanisms, our framework also mitigates adversarial behaviors, bounding the influence of corrupted gradients and preserving a stable flow of parameter updates.
% Our contributions can be summarized as follows:
Therefore, this paper's contributions are summarized as follows:

\begin{itemize}
\item \textbf{Decentralized QKL with Noise Robustness.} We propose a general decentralized framework for QKL which is found robust to handle variable noise levels across different quantum computing units. In particular, theoretical insights are provided to illustrate the consensus convergence even with highly noisy nodes.
% how highly noisy nodes can still converge to a consensus with the rest of the network.
\item \textbf{Defense Against Adversarial Attacks.} We introduce a clipping-based robust aggregation rule that effectively mitigates common adversarial behaviors (e.g., Gaussian and Sign-flipping attacks) in decentralized QKL systems.
\item \textbf{Comprehensive Evaluation.} Numerical simulations on both synthetic and real-world datasets across different network topologies demonstrate that our framework retains high classification accuracy under depolarizing noise and significant adversarial corruption.
\end{itemize}

The remainder of this paper is organized as follows. Section~\ref{sec:related} discusses related work in quantum kernel methods, decentralized optimization, and adversarial robustness. Section~\ref{sec:methodology} describes our proposed methodology for RDQKL, detailing the noise robustness of decentralized QKL and the extension to attack robustness. In Section~\ref{sec:experiments}, we present empirical results on synthetic and benchmark datasets under various network configurations. Finally, Section~\ref{sec:conclusion} concludes the paper and outlines directions for future research on robust, distributed quantum machine learning (QML).

\section{Related Work}
\label{sec:related}

% Quantum kernel learning has emerged as a powerful paradigm within quantum machine learning, similar to quantum neural networks \cite{schuld2019quantum}. Notable advances in this area include theoretical results on achieving quantum advantage in both training efficiency \cite{liu2021rigorous} and expressivity \cite{huang2021power}, which have inspired various extensions such as the Quantum Tangent Kernel \cite{shirai2024quantum}, QKSAN \cite{zhao2024qksan}, and Quantum Kernel-Based LSTM \cite{hsu2024quantum1}. Early applications span science \cite{chen2024validating} and engineering \cite{hsu2024quantum2}, yet many quantum machine learning studies only partially address hardware-induced noise—a critical factor in the noisy intermediate-scale quantum (NISQ) era \cite{cheng2023noisy}. Such noise can exacerbate barren plateaus \cite{larocca2025barren} and reduce the effective dimensionality of quantum models \cite{abbas2021power}, directly impacting generalization performance. Specifically for quantum kernel learning, prior works have shown how Pauli noise leads to exponential concentration \cite{thanasilp2024exponential} and how depolarizing noise degrades predictive accuracy under limited generalization error \cite{wang2025power}. While these studies illuminate important end-point metrics, they devote less attention to the ongoing learning process, such as parameter optimization or kernel alignment \cite{gentinetta2024complexity}.

As a QML paradigm \cite{schuld2019quantum}, similar to quantum neural networks, quantum kernel learning has demonstrated the potential to achieve quantum advantage in terms of both training speed~\cite{liu2021rigorous} and expressivity~\cite{huang2021power}. This has led to the development of various extensions such as Quantum Tangent Kernel~\cite{shirai2024quantum}, QKSAN~\cite{zhao2024qksan}, and Quantum Kernel-Based LSTM~\cite{hsu2024quantum1}. Initial applications are also been seen in fields such as science\cite{chen2024validating} and engineering\cite{hsu2024quantum2}. However, most existing research in QML neglects the impact of noise during analysis and numerical simulations~\cite{cerezo2022challenges}. Actually, in the current era of noisy intermediate-scale quantum (NISQ) devices, hardware-induced noise is a defining characteristic of quantum computation~\cite{cheng2023noisy}. Noise can lead to barren plateaus, thus hampering the trainability of QML models~\cite{larocca2025barren}, and may also reduce the effective dimensionality of the model, thereby impairing its generalization ability~\cite{abbas2021power}.

In the case of quantum kernel learning, prior works have explored the impact of Pauli noise on trainability from the perspective of exponential concentration~\cite{thanasilp2024exponential}, and the effect of depolarizing noise on predictive performance under the assumption of small generalization error~\cite{wang2025power}. These studies focus mainly on the effect of noise on learning outcomes, with limited attention paid to the learning process itself. In fact, when addressing process-related questions—such as how to accelerate quantum kernel alignment~\cite{gentinetta2024complexity}, a method for optimizing parameterized quantum kernels—it becomes especially important to study how noise affects the learning process of quantum kernel learning.

% Beyond noise-related issues, quantum kernel methods face increasing computational challenges as the dimensionality of quantum feature spaces grows. Techniques such as stochastic gradient strategies \cite{gentinetta2023quantum} and sub-sampling \cite{sahin2024efficient} partially alleviate these burdens, but decentralized approaches have recently garnered attention for scaling quantum kernel learning in more realistic settings \cite{ma2024cdqkl}. Decentralization not only aligns with the proliferation of quantum devices but also helps protect sensitive data by minimizing large-scale transfers \cite{chen2024consensus}. Unlike federated learning, which relies on a central aggregator (manager node), decentralized learning distributes coordination tasks among all nodes \cite{qiao2024transitioning, xia2021defending, wu2023byzantine}, alleviating communication bottlenecks and mitigating catastrophic failure risks \cite{liu2019distributed}. Although some works have observed that distributing quantum circuits can enhance noise resilience—by, for instance, limiting circuit depth \cite{avron2021quantum, yu2024quantum}—the effects of hardware noise on decentralized quantum kernel learning (including convergence rates and final accuracies) remain insufficiently explored.

In addition to techniques such as stochastic gradient methods~\cite{gentinetta2023quantum} and sub-sampling~\cite{sahin2024efficient}, decentralized approaches have been proposed to further address the computational challenges in quantum kernel learning~\cite{ma2024cdqkl}. In fact, decentralized methods not only align with the trend of increasingly ubiquitous quantum devices, but also help mitigate the risk of leaking private or sensitive data~\cite{chen2024consensus}. Unlike federated learning~\cite{qiao2024transitioning,xia2021defending}, decentralized learning has emerged as a simple, effective, and more general approach~\cite{wu2023byzantine}.  Compared to the potential issues of communication bottlenecks on the manager node that may lead to the failure of all connected machines in the case of federated learning with a manager-worker architecture, decentralized learning network could avoid potential failures and communication latency~\cite{liu2019distributed}. Although prior studies have explored the noise-resilience benefits of distributed quantum computation from the perspective of shallower-depth circuits~\cite{avron2021quantum,yu2024quantum}, the direct influence of noise on the decentralized method has not yet been investigated.

% As quantum computing moves closer to practical use, security and adversarial threats also come to the fore \cite{mi2022securing}. Model extraction attacks can compromise quantum models by exposing proprietary circuits or parameters \cite{kundu2024evaluating}, backdoor attacks can enable targeted false predictions \cite{chu2023qdoor}, and data poisoning can corrupt training inputs within quantum environments \cite{kundu2024adversarial}. In a decentralized quantum machine learning setup, these vulnerabilities may intensify due to an expanded pool of participants and the absence of a single controlling authority \cite{wu2023byzantine}. Although an auction-based mechanism has been proposed to filter out untrustworthy nodes \cite{lee2025auction}, its high deployment cost limits broader applicability. Thus, there is a pressing need for simpler, more generally applicable methods to defend decentralized quantum learning frameworks against both noise and malicious behaviors \cite{he2022byzantine}.

As quantum computing progresses toward practical applications, it also introduces new security challenges~\cite{mi2022securing}. Researchers have begun to examine the effectiveness of model extraction attacks on QML models~\cite{kundu2024evaluating}. Another study has highlighted the threat of backdoor attacks targeting QML models~\cite{chu2023qdoor}, while data poisoning attacks have also been explored in the context of QML environments~\cite{kundu2024adversarial}. Although decentralized QML offers numerous advantages, its decentralized communication structure and large number of participants pose even greater security risks~\cite{wu2023byzantine}. Recently, one work has introduced an auction-based mechanism to filter out unreliable clients~\cite{lee2025auction}. However, this increases the deployment cost of decentralized QML. There is a need for a more simpler and general strategy to enhance the robustness of decentralized methods~\cite{he2022byzantine}.

\section{ PRELIMINARIES}

This section establishes the mathematical and physical foundations of the quantum kernel methods used throughout this work. We begin by recalling the structure of quantum states and operators, then introduce the concept of kernel functions in quantum feature spaces. Finally, we address how noise---especially depolarizing errors---affects the fidelity of quantum kernels and discuss the implications for practical machine learning tasks.

\subsection{Quantum States and Operators}

\subsubsection{1. Hilbert Space and Qubits}

A quantum system of \(N\) qubits is described by a complex Hilbert space \(\mathcal{H} \cong \mathbb{C}^{2^N}\). Any pure state \(|\psi\rangle\) in \(\mathcal{H}\) is a normalized vector, \(\|\psi\| = 1\). However, due to decoherence and interactions with the environment, a general description of the system may require the concept of a density operator \(\rho\). Formally, \(\rho\) is a positive semidefinite, Hermitian operator of trace one:
\begin{equation}
\rho \succeq 0, \quad \rho = \rho^\dagger, \quad \mathrm{Tr}(\rho) = 1.
\end{equation}
A density operator with \(\mathrm{rank}(\rho)=1\) corresponds to a pure state \(\rho = |\psi\rangle\langle \psi|\). States with higher rank are known as mixed states and can be interpreted as probabilistic ensembles of pure states.

\subsubsection*{2. Measurements and Observables}

Observables in quantum mechanics are represented by Hermitian operators \(O\) acting on \(\mathcal{H}\). Measuring \(\rho\) with respect to \(O\) yields an expectation value \(\mathrm{Tr}(O\,\rho)\). When making binary decisions (e.g., labeling data as \(+1\) or \(-1\)), one often uses projective or POVM (Positive Operator-Valued Measure) measurements. In learning contexts, one might designate a measurement operator \(O\) to map a quantum state to a real-valued label:
\begin{equation}
y := \mathrm{Tr}\bigl(O\,\rho\bigr), \quad \|O\|_2 \le 1,
\end{equation}
where \(\|\cdot\|_2\) denotes the spectral norm, bounding the maximal eigenvalue of \(O\).

\subsection{Quantum Kernel Methods}

Classical kernel methods (e.g., support vector machines, kernel ridge regression) rely on feature maps \(\phi: X \to F\) from an input space \(X\) into a high-dimensional feature space \(F\). A quantum kernel method adopts a quantum feature map, using an \(N\)-qubit circuit that embeds data into \(\mathcal{H}\) and leverages quantum measurement to compute inner products.

\subsubsection*{1. Quantum Feature Map}

Let \(x \in X \subset \mathbb{R}^d\) be a classical data vector. A parameterized unitary operator
\begin{equation}
U_E(\mathbf{x}): |0\rangle^{\otimes N} \mapsto |\phi(\mathbf{x})\rangle \in \mathcal{H}
\end{equation}
encodes \(x\) into a quantum state. The resulting density operator is
\begin{equation}
\rho(\mathbf{x}) = |\phi(\mathbf{x})\rangle\langle \phi(\mathbf{x})|.
\end{equation}
Because \(|\phi(\mathbf{x})\rangle\) may occupy a large Hilbert space (dimension \(2^N\)), quantum feature maps are hypothesized to capture complex relationships among data that are difficult to replicate classically.

\subsubsection*{2. Kernel Function via State Overlap}

The quantum kernel function between two samples \(x\) and \(x'\) is defined as
\begin{equation}
K(\mathbf{x}, \mathbf{x}') = \mathrm{Tr}\bigl[\rho(\mathbf{x})\,\rho(\mathbf{x}')\bigr] = \bigl|\langle \phi(\mathbf{x}) \mid \phi(\mathbf{x}')\rangle\bigr|^2.
\label{eq:qkernel_def}
\end{equation}
This overlap can be experimentally estimated by building an interference circuit that effectively compares \(|\phi(\mathbf{x})\rangle\) and \(|\phi(\mathbf{x}')\rangle\). For each pair \(x, x'\) in the training set, one obtains an empirical estimate \(\widehat{K}(x,x')\). Collecting these values into a matrix \(K\), we arrive at an \(n \times n\) Gram matrix of quantum-induced inner products.

\subsubsection*{3. Classical Optimization on Quantum Kernels}

Once \(K\) is available, classical machine learning techniques (SVM, Gaussian processes, kernel ridge regression, etc.) handle the final optimization. For instance, in a regularized regression approach with labels \(\{y_i\}\),
\begin{equation}
\omega^* = \underset{\omega}{\mathrm{argmin}} \, \sum_{i=1}^n \Bigl( y_i - \mathrm{Tr}\bigl[\rho(x_i)\,\omega\bigr] \Bigr)^2 + \lambda\,\|\omega\|^2,
\end{equation}
where \(\omega\) can be represented in the span of \(\{\rho(x_i)\}\). Equivalently, in the dual formulation, \(\omega^*\) is related to \((K + \lambda I)^{-1}\mathbf{y}\). Thus, the model prediction for a new \(x\) is given by
\begin{equation}
h(x) = \mathrm{sign}\Bigl[\mathrm{Tr}\bigl(\rho(x)\,\omega^*\bigr)\Bigr] = \mathrm{sign}\Bigl[K(x,\cdot)\,(K + \lambda I)^{-1}\,\mathbf{y}\Bigr],
\end{equation}
up to saturation bounds (e.g., \(\pm 1\) for classification).

\subsection{Noise in Quantum Kernels}

\subsubsection*{1. General Framework of Quantum Noise}

In realistic quantum devices, perfect unitaries are hard to implement---especially in the Noisy Intermediate-Scale Quantum (NISQ) setting. A quantum channel \(\mathcal{E}\) representing noise can be described by a set of Kraus operators \(\{E_\alpha\}\) such that
\begin{equation}
\mathcal{E}(\rho) = \sum_\alpha E_\alpha\,\rho\, E_\alpha^\dagger, \quad \sum_\alpha E_\alpha^\dagger E_\alpha = I.
\end{equation}
When one composes multiple quantum gates or layers, noise typically accumulates, altering the output state and thus the measured kernel value.

\subsubsection*{2. Depolarizing Noise Model}

A common noise model in quantum computing is depolarizing noise, which replaces the input state \(\rho\) with the maximally mixed state \(I/D\) (where \(D = 2^N\) is the dimension) at a certain probability rate. Formally, for a single-layer depolarizing channel with parameter \(\tilde{p}\):
\begin{equation}
\mathcal{N}_{\tilde{p}}(\rho) = (1 - \tilde{p})\,\rho + \tilde{p}\,\frac{I}{D}.
\label{eq:one_layer_depol}
\end{equation}
If an encoding circuit of depth \(L\) is subject to noise \(\mathcal{N}_{\tilde{p}}\) after every gate layer, then the effective global depolarizing rate across the entire circuit becomes
\begin{equation}
p = 1 - (1 - \tilde{p})^{2L},
\end{equation}
assuming \(2L\) layers in total (e.g., forward and inverse layers in certain kernel measurement schemes) \cite{wang2025power}.

\begin{figure*}[htbp]
\begin{center}
\centerline{\includegraphics[width=0.9\textwidth]{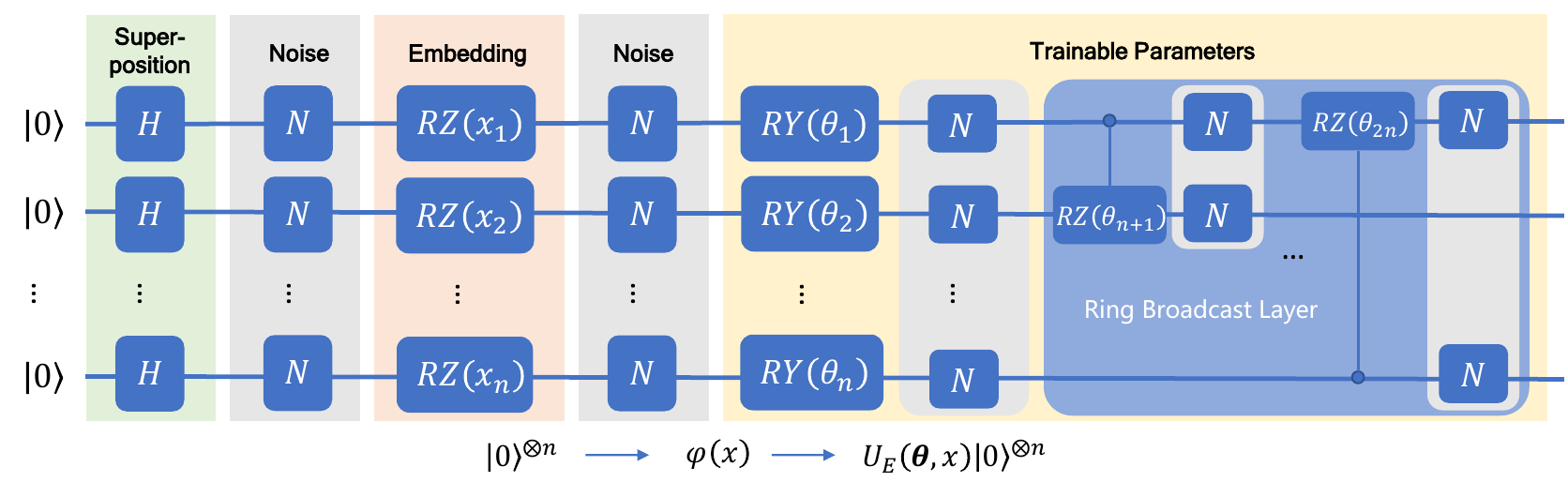}}
\caption{A quantum feature mapping circuit under noise with variational parameters. The circuit can be sequentially divided into a superposition layer, an embedding layer and a trainable parameter layer. The superposition layer applies Hardman gate to each qubit to build a superposition state. The embedding layer applies a z-axis rotation with an angle of $x_i$ to the $i^{th}$ qubit thereby embedding the information of $x$ into the working system. The trainable parameter layer includes the y-axis rotation applied to each qubit and a ring broadcast layer in order to create entanglement and broadcast information. Each quantum gate is followed by a depolarized noise channel.}
% \vskip -0.1in
\label{fig_c}
\end{center}
\end{figure*}

\subsubsection*{3. Noisy Quantum Kernel}

Let \(\rho_0(x, x')\) be the ideal output of the interference circuit that compares \(|\phi(x)\rangle\) and \(|\phi(x')\rangle\). In the absence of noise,
\begin{equation}
\rho_0(x, x') = |\phi(x, x')\rangle \langle \phi(x, x')|.
\end{equation}
Under global depolarizing noise \(\mathcal{N}_p\), the state transforms as
\begin{equation}
\rho(x, x') = \mathcal{N}_p\bigl[\rho_0(x, x')\bigr] = (1-p)\,\rho_0(x, x') + p\,\frac{I}{D}.
\end{equation}
Hence, the noisy kernel \(K_e\) measured by projecting onto \(|0\rangle^{\otimes N}\) (or an equivalent basis projector \(P_0\)) becomes
\begin{align}
K_e(x, x') &= \mathrm{Tr}\Bigl[P_0\,\rho(x, x')\Bigr] \nonumber\\[1mm]
&= (1-p)\,\mathrm{Tr}\bigl[P_0\,\rho_0(x, x')\bigr] + p\,\frac{1}{D}.
\label{eq:noisy_kernel}
\end{align}
If \(\mathrm{Tr}[P_0\,\rho_0(x, x')]\) is the ideal kernel \(K(x, x')\), then
\begin{equation}
K_e(x, x') = (1-p)\,K(x, x') + p\,\frac{1}{D}.
\label{14}
\end{equation}
Notably, in the extreme limit \(p \to 1\), the kernel value collapses to a constant \(1/D\), rendering the classifier (or regressor) incapable of distinguishing distinct inputs.

\subsubsection*{4. Measurement Statistics in Practice}

In an experimental scenario, each kernel value \(K_e(x, x')\) must be sampled via repeated measurements (shots). Let \(m\) denote the number of repeated measurements to estimate one kernel entry. Each shot may be viewed as a Bernoulli trial with success probability \(K_e(x, x')\). The empirical estimate
\begin{equation}
\widehat{K}_e(x, x') = \frac{1}{m}\sum_{k=1}^m V_k(x, x'),
\end{equation}
where each \(V_k \in \{0,1\}\) with expectation \(\mathbb{E}[V_k] = K_e(x, x')\). By the law of large numbers and standard concentration inequalities (e.g., Hoeffding’s inequality), for sufficiently large \(m\), \(\widehat{K}_e\) converges to \(K_e\) with high probability. In this paper, we assume that \(m\) is large enough.

\subsubsection*{5. Consequences for Model Training}

In a kernel-based learning algorithm, one replaces the ideal kernel matrix \(K\) with its noisy (and possibly shot-based) counterpart, \(K_e\). The final hypothesis for a new point \(x\) becomes
\begin{equation}
\hat{h}(x) = \mathrm{sign}\Bigl[\bigl(K_e(x,\cdot)\bigr)^\top (K_e + \lambda I)^{-1} y\Bigr],
\label{eq:noisy_hypothesis}
\end{equation}
up to any clipping or thresholding to ensure labels stay in \([-1,1]\) or a desired range. When \(p\) is small and \(m\) is large, \(K_e \approx K\), enabling effective learning. As \(p\) grows or \(m\) shrinks, the noise dominates, and performance degrades.

\section{Methodology}
\label{sec:methodology}
\subsection{Quantum Kernel Learning}

In order to improve the quantum kernel's performance on a specific dataset $D_0$, quantum kernel alignment can be used to train the variational quantum circuit $U_E(\bm{\theta}, x)$ just like the circuit shown in Fig.\ref{fig_c} for quantum feature mapping, such that $K_e(\bm{\theta}, x_i, x_j)$ approaches an ideal kernel function, which can be expressed as
\begin{equation}
    K^*(\bm\theta^*, x_i, x_j) = y_i y_j.
\end{equation}

This implies the output is $1$ when $x_i$ and $x_j$ belong to the same class and $-1$ when they belong to different classes.The alignment value measures the closeness between $K_e(\bm{\theta})$ and $K^*$, and is expressed as
\begin{equation}
\begin{aligned}
    A(K_e(\bm{\theta}), K^*) &= \frac{\langle K_e(\bm{\theta}), K^* \rangle_F}{\sqrt{\langle K_e(\bm{\theta}), K_e(\bm{\theta}) \rangle_F \langle K^*, K^* \rangle_F}} \\&= \frac{\sum_{i,j} y_i y_j K_e(\bm{\theta}, x_i, x_j)}{n \sqrt{\sum_{i,j} K_e^2(\bm{\theta}, x_i, x_j)}},
\end{aligned}
\end{equation}
where $\langle \cdot, \cdot \rangle_F$ denotes the Frobenius inner product of two matrices. Intuitively, when the labels of data points are correctly predicted, $A(K_e(\bm{\theta}), K^*)$ will increase, and when the labels of data points are incorrectly predicted, $A(K_e(\bm{\theta}), K^*)$ will decrease. If we define an objective function as
\begin{equation}
    \min_{\theta \in \mathbb{R}^T} L(D_0, \bm{\theta}) = -A(K_e(\bm{\theta}), K^*).
\end{equation}

Upon obtaining gradients through methods like parameter shift, the gradient descent method can be employed to iteratively update the parameter $\bm{\theta}$ towards the optimal value $\bm\theta^*$, thereby aligning $K_e(\bm{\theta})$ with the ideal kernel $K^*$.

\subsection{Effect of Noise on Quantum Kernel Learning}

The presence of noise can degrade the discriminative power of quantum kernels. Existing study has shown that, for a given number of training samples, once the noise intensity or the number of noise-affected layers exceeds a certain threshold, the discriminative capability of the noisy kernel deteriorates significantly \cite{wang2025power}. This implies that when the noise impact is strong, the effectiveness of QKL is reduced. In other words, even after prolonged quantum kernel training, the discriminative power of the quantum kernel remains poor.

\begin{figure}[t]
\begin{center}
\centerline{\includegraphics[width=\columnwidth]{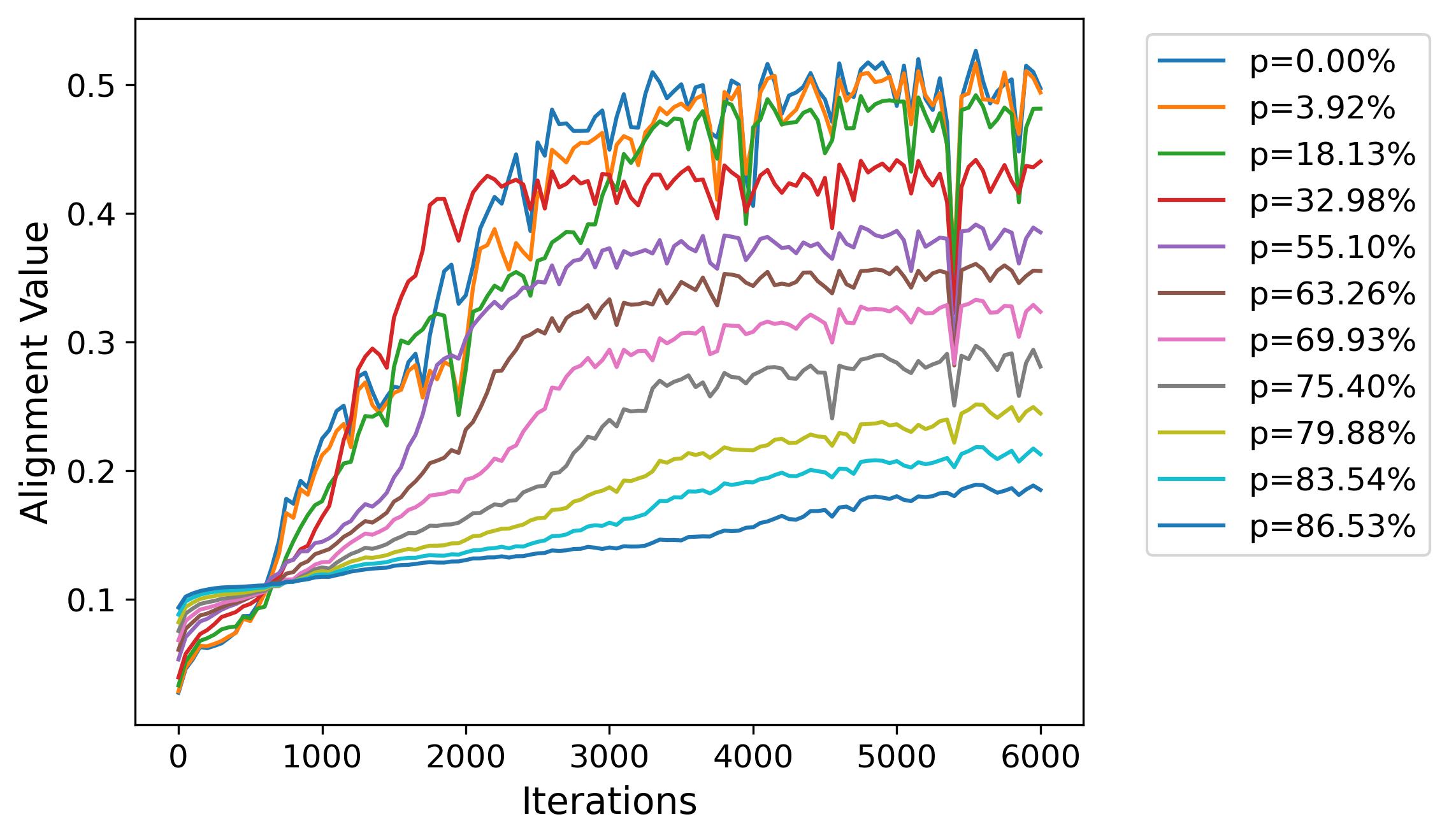}}
\caption{Quantum kernel alignment value  during the training process under different noise intensities. The larger the quantum kernel alignment value, the better the quantum kernel distinguishes on the specific data set. The quantun circuit is described in A of Section \ref{sec:experiments}. }
% \vskip -0.2in
\label{fig1}
\end{center}
\end{figure}

Moreover, the presence of noise will affect the  process of QKL. Here we have a proposition. When the noise is not significant, its presence may either accelerate or decelerate the QKL. However, when the noise becomes considerably large, the learning process is generally slowed down. This phenomenon can be visually observed in a example as shown in the fig. \ref{fig1}. When the equivalent noise intensity is $32.98\%$, the training speed is actually higher than that under lower noise levels. As the noise continues to increase, the speed of the QKL gradually decreases.

% Moreover, the presence of noise will affect the speed of quantum kernel learning. As illustrated in the Fig. \ref{fig1}, the process of quantum kernel learning slows down as noise increases. From a theoretical perspective, according to Eq. \ref{14}, a noisy quantum kernel can be regarded as a linear scaling of the ideal quantum kernel. Consequently, to achieve a specific quantum kernel value $K_e(\bm{\theta}) = K_0$, more training iterations are required to obtain parameters $\bm{\theta}_0$ such that $K(\bm{\theta}_0, x_i, x_j)$ reaches a value that is either smaller or larger than $K_0$, specifically $(K_0 - p/D)/(1 - p)$, rather than $K_0$. 

An explanation can be given from the perspective of gradient. In the absence of noise, the gradient of the quantum kernel alignment value with respect to $\theta_t$ can be expressed as

\begin{equation}
\begin{aligned}
\frac{\partial A(K(\bm{\theta}), K^*)}{\partial \theta_t} &= 
\frac{\sum_{ij} y_i y_j \frac{\partial K_{ij}(\bm{\theta})}{\partial \theta_t}}{n \sqrt{\sum_{ij} K_{ij}^2 (\bm{\theta})}}
-\\&
\frac{\left( \sum_{ij} y_i y_j K_{ij}(\bm{\theta}) \right) \sum_{ij} K_{ij}(\bm{\theta}) \frac{\partial K_{ij}(\bm{\theta})}{\partial \theta_t}}
{n \left( \sum_{ij} K_{ij}^2 (\bm{\theta}) \right)^{3/2}}.
\end{aligned}
\end{equation}

When noise is present, the quantum kernel function is given by  $K_{e_{ij}} (\bm{\theta} ) = (1 - p) K_{ij}(\bm{\theta} ) + \frac{p}{D}$.
Assuming the dataset is balanced, the gradient of the quantum kernel alignment value with respect to $\theta_t$ is given by

\begin{equation}
\begin{aligned}
&\frac{\partial A(K_e (\bm{\theta}), K^*)}{\partial \theta_t} = 
\frac{\sum_{ij} y_i y_j \frac{\partial K_{ij} (\bm{\theta})}{\partial \theta_t}}
{n \sqrt{\sum_{ij} \left(K_{ij} (\bm{\theta}) + \frac{p}{(1 - p) D} \right)^2 }}
\\&- 
\frac{\left( \sum_{ij} y_i y_j K_{ij} (\bm{\theta}) \right) \sum_{ij} \left(K_{ij} (\bm{\theta}) + \frac{p}{(1 - p) D} \right) \frac{\partial K_{ij} (\bm{\theta})}{\partial \theta_t}}
{n \left( \sum_{ij} \left(K_{ij} (\bm{\theta}) + \frac{p}{(1 - p) D} \right)^2 \right)^{3/2} }
\end{aligned}
\end{equation}

When the noise intensity is relatively low, the magnitude of the gradient with noise, \(\partial A(K_e (\bm{\theta}), K^*)/{\partial \theta_t}\), is indistinguishable from the noiseless gradient, $\partial A(K (\bm{\theta}), K^*)/\partial \theta_t$. In this situation, noise essentially introduces additional randomness into the training process, which may either accelerate or decelerate the learning of the quantum kernel \cite{liu2022stochastic}.

Assuming that $\partial K_{ij}(\bm{\theta} )/\partial \theta_t$ is bounded, when the noise intensity is relatively high, the gradient under noise \(\partial A(K_e (\bm{\theta}), K^*)/{\partial \theta_t}\), will be smaller than the gradient without noise $\partial A(K (\bm{\theta}), K^*)/\partial \theta_t$, due to the increase in the denominator. This implies that when the noise influence is strong, the process of the QKL using fixed step size will slow down. Furthermore, as \( p \) approaches $1$, $\partial A(K_e (\bm{\theta}), K^*)/\partial \theta_t$ will approaches $0$. In other words, when the noise intensity is very high, the gradient of the quantum kernel alignment with respect to the parameters becomes very small.

\subsection{General Decentralized Quantum Kernel Learning}

We consider a decentralized network described by an undirected connected graph $G(V, E)$, where $V = \{1, \dots, N\}$ represents the quantum computing units in the network, and $E \subseteq V \times V$ denotes the connections between quantum computing units via classical channels. In this setting, each quantum computing unit has a local dataset $D_i$, and the entire network's dataset is defined as  $D \triangleq \bigcup_{i=1}^{N} D_i$.  The objective of the network is to minimize the loss function over the entire dataset, given by
\begin{equation}
 \begin{aligned}
    \min_{\bm{\theta} \in \mathbb{R}^T} L(D, \bm{\theta}) &=  \sum_{i=1}^{N} L_i(D_i, \bm{\theta})\\&= -\sum_{i=1}^{N} A_i(K_{e_i}(\bm{\theta}), K_i^*).
\end{aligned}
\end{equation}

In this decentralized network, to obtain the optimal parameter $\bm{\theta}^*$ that minimizes $L(D, \bm{\theta})$, a decentralized approach, as illustrated in \textbf{Algorithm} \ref{alg1}, can be employed.  Overall, the method consists of three main components: local gradient descent, data communication between nodes, and model information aggregation.

First, at time step $k$, each quantum computing unit $i \in V$ performs a gradient descent update with step size $\eta_i$ based on its local dataset to update the parameters of the variational quantum circuit from $\bm{\theta}_i^k$ to $\bm{\theta}_i^{(k+\frac{1}{2})}$. That is
\begin{equation}
    \bm{\theta}_i^{(k+1/2)} = \bm{\theta}_i^k - \eta_i \nabla_{\bm{\theta}} L_i(D_i, \bm{\theta}_i^k).
\end{equation}
Here, a subsampling strategy can be employed, where the stochastic gradient of the loss function over $D_i$ is approximated using $q_i$ randomly selected samples from $D_i$. The stochastic gradient of the loss function is given by
\begin{equation}
    \tilde{\nabla}_{\bm{\theta}} L_i (D_i, \bm{\theta}_i^k) = \frac{1}{q_i} \sum_{p=1}^{q_i} \nabla_{\bm{\bm{\theta}}} L_p (s_i^p, \bm{\theta}_i^k).
\end{equation}

Next is the stage of data communication between nodes. If $(i, j) \in E$, then quantum computing unit $j$ is considered a neighbor of quantum computing unit $i$, and we define the neighbor set as $N_i \triangleq \{ j \mid (i, j) \in E, \forall j \in V \} \cup \{ i \}$. At this stage, quantum computing unit $i$ sends $\bm{\theta}_i^{(k+\frac{1}{2})}$ to all $j \in N_i$ and simultaneously receives $\bm{\theta}_{(i,j)}^{(k+\frac{1}{2})}$ from all $j \in N_i$.

Finally, each quantum computing unit $i \in V$ will compute the final model parameters at time step $k$ as
\begin{equation}
    \bm{\theta}_i^{(k+1)} = A_i \left( \left\{\bm{\theta}_{(i,j)}^{(k+\frac{1}{2})} \right\}_{j \in N_i} \right),
\end{equation}
where the aggregation rule $A_i: \mathbb{R}^{T \times |N_i|} \to \mathbb{R}^T$ is applied. A common aggregation strategy is weighted averaging. If we define the weight matrix $W = [w_{ij}]_{N \times N}$, where $w_{ij} > 0$ if $(i,j) \in E$ and $w_{ij} = 0$ otherwise, and $W$ satisfies $W \mathbf{1}_N = \mathbf{1}_N$ and $\mathbf{1}_N^T W = \mathbf{1}_N^T$, meaning that $W$ is a doubly stochastic matrix, then the aggregation rule $A_i$ can be expressed as:  
\begin{equation}
    A_i \left( \left\{ \bm{\theta}_{(i,j)}^{(k+\frac{1}{2})} \right\}_{j \in N_i} \right) = \sum_{j \in N_i} w_{ij} \bm{\theta}_{(i,j)}^{(k+\frac{1}{2})}.
\end{equation}

\begin{algorithm}[tb]
   \caption{General Decentralized Quantum Kernel Learning }
   \label{alg1}
\begin{algorithmic}[1]
   \STATE {\bfseries Input:} sub-dataset $D_i$ for $i$-th quantum computing unit
   \FOR{$\forall i \in V$}
    \STATE Establish connections with quantum neighbors in $N_i$
    \STATE Initialize the the aggregation rule $A_i$
    \STATE Initialize quantum parameters $\bm{\theta}_i^0$ and training step $\eta_i$
   \ENDFOR
   \REPEAT
   \FOR{$\forall i \in V$}
   \STATE Gradient descent: $\bm{\theta}_i^{(k+\frac{1}{2})} = \bm{\theta}_i^k - \eta_i  \tilde{\nabla}_{\bm{\theta}} L_i (D_i, \bm{\theta}_i^k)$
   \STATE Data exchange:  send $\bm{\theta}_i^{(k+\frac{1}{2})}$ to $\forall j \in N_i$ and simultaneously receives $\bm{\theta}_{(i,j)}^{(k+\frac{1}{2})}$ from $\forall j \in N_i$
   \STATE Model aggregation: $\bm{\theta}_i^{(k+1)} = A_i \left( \left\{ \bm{\theta}_{(i,j)}^{(k+\frac{1}{2})} \right\}_{j \in N_i} \right)$
   \ENDFOR
   \STATE Set $k:=k+1$
   \UNTIL{Termination condition is satisfied  such as the norm of convergence gradient $\|g\|< g_{thresh}$ or iteration budget is approached and obtain $\bm{\theta}_i^e$ for $\forall i \in V$}
\end{algorithmic}
\end{algorithm}

\subsection{Noise Resilience of Decentralized QKL}

In the presence of noise, if the noise of individual quantum computing units is significantly higher, decentralized QKL can still achieve effective training and reach consensus on each node. The following will take the weighted average aggregation method as an example to illustrate.

Regarding the effectiveness of training, it can be analyzed by observing the evolution of the average quantum variational parameters of all quantum computing units, given by  
\begin{equation}
    \bar{\bm{\theta}}^k = \frac{1}{N} \sum_{i=1}^{N} \bm{\theta}_i^k = \frac{1}{N} \mathbf{1}_N^T \bm{\theta}^k,
\end{equation}
over the time step $k$. To this end, we consider the matrix representation of the weighted average aggregation method,
\begin{equation}
    \bm{\theta}^{k+1} = W\bm{\theta}^k - \eta \cdot W \nabla \bm{F}(\bm{\theta}^k),
\end{equation}
where $\nabla \bm{F}(\bm{\theta}^k)$ is defined as the pseudo gradient of $L(D, \bm{\theta})$, expressed as
\[
\nabla \bm{F}(\bm{\theta}^k) \triangleq 
\begin{bmatrix} 
\tilde{\nabla}_{\bm{\theta}} L_1 (D_1, \bm{\theta}_1^k) \\ 
\tilde{\nabla}_{\bm{\theta}} L_2 (D_2, \bm{\theta}_2^k) \\ 
\vdots \\ 
\tilde{\nabla}_{\bm{\theta}} L_N (D_N, \bm{\theta}_N^k) 
\end{bmatrix}.
\]
Thus, the average over all quantum computing units at time step $k+1$ can be written as
\begin{equation}
\begin{aligned}
    \bar{\bm{\theta}}^{(k+1)} &= \frac{1}{N} \mathbf{1}_N^T \bm{\theta}^{(k+1)}
\\&= \frac{1}{N} \left( \mathbf{1}_N^T W \bm{\theta}^k - \eta \cdot \mathbf{1}_N^T W \nabla\bm{F}(\bm{\theta}^k) \right).
\end{aligned}
\end{equation}
Since \( W \) is a doubly stochastic matrix, we have \( \mathbf{1}_N^T W = \mathbf{1}_N^T \), which leads to
\begin{equation}
    \begin{aligned}
        \bar{\bm{\theta}}^{(k+1)} &= \frac{1}{N} \left( \mathbf{1}_N^T \bm{\theta}^k - \eta \cdot \mathbf{1}_N^T \nabla \bm{F}(\bm{\theta}^k) \right)
\\&= \bar{\bm{\theta}}^k - \eta \cdot \tilde{\nabla}_{\bm{\theta}} L (D, \bm{\theta}^k),
    \end{aligned}
\end{equation}
where $\nabla_{\bm{\theta}} \tilde{L}(D, \bm{\theta}^k) \triangleq \frac{1}{N} \mathbf{1}_N^T \nabla F(\bm{\theta}^k)$ can be regarded as the stochastic gradient of $L(D, \bm{\theta})$. Therefore, decentralized QKL can be viewed as a stochastic gradient descent method applied to the mean value of the quantum variational parameters.

When certain quantum computing units experience relatively high noise levels, other low-noise quantum computing units can provide more accurate gradient information, thereby accelerating their QKL process. Meanwhile, the learning information from these higher noisy units can still be leveraged by other low-noise quantum computing units.

When the noise intensity of certain quantum computing units becomes excessively high, the local gradient of the quantum kernel alignment value with respect to the parameters approaches zero. Consequently, the higher noisy units will not significantly affect other low-noise quantum computing units.

Regarding consensus, it can be analyzed by
\begin{equation}
\begin{aligned}
    \bm{\theta}^{(k+1)} &= W\bm{\theta}^k - \eta W \nabla \bm{F}(\bm{\theta}^k) \\&= W^k \bm{\theta}^0 - \eta \sum_{0 \leq i \leq k} W^{(k-i)} \nabla \bm{F}(\bm{\theta}^i).
\end{aligned}
\end{equation}

Since $W$ is a doubly stochastic matrix, under the condition
\begin{equation}
    \rho(W - \frac{1_N 1_N^T}{N}) < 1,
\end{equation}
$W^k$ will converge to $\frac{1_N 1_N^T}{N}$ as $k \to \infty$. If $\nabla F(\theta^i)$ is bounded and converges to zero, then $\theta_i^{\infty}$ will converge to a common value $\theta^{\infty}$ \cite{kia2019tutorial}. This indicates that even when certain quantum computing units experience significantly higher noise levels, they will not hinder the consensus among all quantum computing units.

\subsection{Robust Strategy Against Adversarial Information Attack}

Decentralized quantum kernel learning involves a large number of quantum computing units. If data corruption, device failures, or malicious attacks occur, the system may deviate from the training protocol. Here, we consider two types of information attacks, Gaussian attack and Sign-flipping attack.

During the data exchange stage of \textbf{Algorithm} \ref{alg1}, an attacker $i$ first collects information $\theta_j^{(k+\frac{1}{2})}$ from all neighboring nodes $\forall j \in N_i$. Based on collected information, the attacker then generates a attack value $\theta_i^{(k+\frac{1}{2})}$, which is subsequently sent to all neighbors $\forall j \in N_i$.

For Gaussian attack, the attack value $\theta_i^{(k+\frac{1}{2})}$ is computed as
\begin{equation}
\theta_i^{(k+\frac{1}{2})} = \text{Gaussian} \big(\text{Average}(\theta_j^{(k+\frac{1}{2})}), \text{Variance}(\theta_j^{(k+\frac{1}{2})}) \big),
\end{equation}
where $\text{Average}(\cdot)$ and $\text{Variance}(\cdot)$ represent the mean and variance functions respectively and $\text{Gaussian}(a,b)$ generates a Gaussian random variable with mean $a$ and variance $b$.

For Sign-flipping attack, the attack value $\theta_i^{(k+\frac{1}{2})}$ is computed as
\begin{equation}
\theta_i^{(k+\frac{1}{2})} = -\text{Average}(\theta_j^{(k+\frac{1}{2})}).
\end{equation}

Intuitively, in an untrusted environment, although the received data $\theta_{(i,j)}^{(k+\frac{1}{2})}$ from neighboring quantum computing units may be unreliable, the local data $\theta_i^{(k+\frac{1}{2})}$ is considered trustworthy. Therefore, it can be used as a reference to evaluate and clip the received data from neighbors. A clipping method $\text{Clip}(\theta_{(i,j)}^{(k+\frac{1}{2})}, \tau)$ can be expressed as
\begin{equation}
\text{Clip}(\theta_{(i,j)}^{(k+\frac{1}{2})}, \tau) = \min\left(1, \frac{\tau}{\|\theta_{(i,j)}^{(k+\frac{1}{2})}\|} \right) \cdot \theta_{(i,j)}^{(k+\frac{1}{2})}
\end{equation}
where $\tau$ is the threshold. A smaller $\tau$ leads to a more sensitive attack detection but may slow down training, while a larger $\tau$ reduces the robustness against attacks and makes the performance closer to that of the general decentralized method.

To mitigate adversarial attacks, the following robust aggregation rule $A_i^r$ can be used:
\begin{equation}
A_i^r \left(\{\theta_{(i,j)}^{(k+\frac{1}{2})} \}_{j \in N_i} \right) = \sum_{j \in N_i} w_{ij} \text{Clip}(\theta_{(i,j)}^{(k+\frac{1}{2})}, \tau).
\end{equation}

If the general aggregation rule $A_i \left( \left\{ \bm{\theta}_{(i,j)}^{(k+\frac{1}{2})} \right\}_{j \in N_i} \right)$ in \textbf{Algorithm} \ref{alg1} is replaced by the robust aggregation rule $A_i^r \left(\{\theta_{(i,j)}^{(k+\frac{1}{2})} \}_{j \in N_i} \right)$, the RDQKL algorithm will be formed.

\section{Experiments}
\label{sec:experiments}
\subsection{Experiment Settings}

\begin{figure}[ht]
\begin{center}
\centerline{\includegraphics[width=\columnwidth]{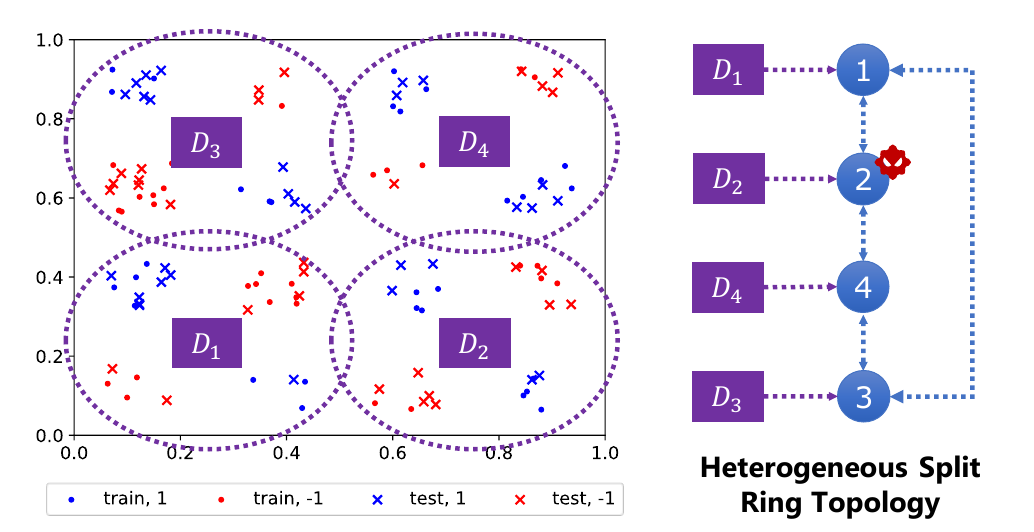}}
\vskip -0.1in
\caption{The checkerboard data is within a 1x1-sized region, divided into a total of 16 checkerboard cells , each of which is a square of $0.25\times0.25$. Each cell contains Gaussian random data centered around the respective checkerboard. The checkerboard dataset is heterogeneously partitioned into a ring topology and Node $2$ is set as a high noise or malicious node.}
\label{data fig1}
\end{center}
\end{figure}

\begin{figure}[ht]
\begin{center}
\centerline{\includegraphics[width=0.9\columnwidth]{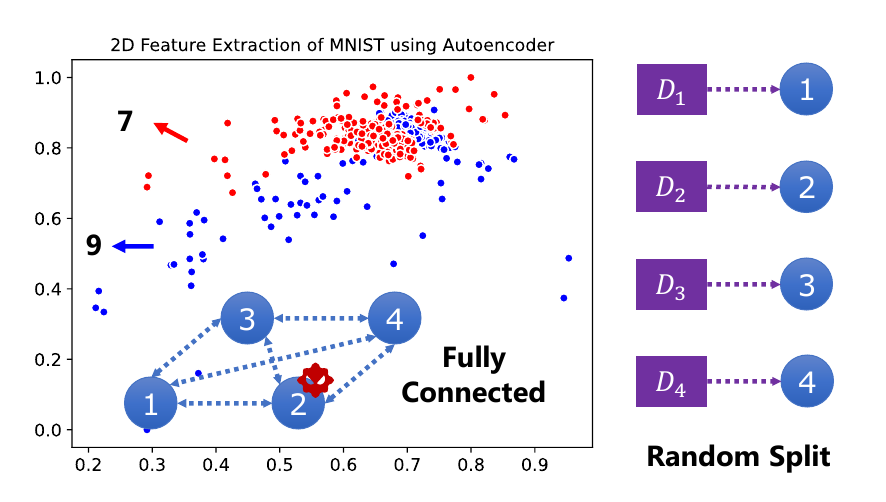}}
\vskip -0.1in
\caption{The MNIST dataset is reduced in dimension to a 1x1-sized region using an autoencoder. The encoder, which contains $4$ fully connected layers with gradually decreasing dimensions, maps them to a 2D representation, and the decoder, which contains 4 corresponding fully connected layers, attempts to reconstruct the original input data for digits 7 and 9 from this low-dimensional representation. The reduced-dimensional MNIST dataset is randomly divided into a fully connected topology and Node $2$ is set as a high noise or malicious node.}
\label{data fig2}
\end{center}
\end{figure}

\begin{table*}[ht]
    \centering
\caption{Experiment results on dataset 1 under noise environment}
\label{tab1}
    \begin{tabular}{cccccc}
    \toprule
    \textbf{Scenario} & \textbf{Experiment Setting} & \textbf{Iteration Times} & \textbf{Score1} & \textbf{Score2} & \textbf{Score3}\\
    \midrule 
        \multirow{3}{*}{\makecell[c]{Normal\\Noise}} & Decentralized QKL w/ $\tilde{p}_{all}=0.05 \%$ & 1050 & 100.00\% (avg.) & 81.67\% (avg.) & 93.33\% (avg.)\\
        &Centralized QKL w/ $\tilde{p}=0.05 \%$ & 3900 & \rule[0pt]{1em}{0.4pt} & \rule[0pt]{1em}{0.4pt} & 98.33\%\\
        &Local QKL /w $\tilde{p}_{all}=0.05 \%$ & 750 & 100.00\% (avg.)& 66.67\% (avg.)& 85.42\% (avg.)\\
    \midrule
        \multirow{4}{*}{\makecell[c]{Relatively\\High Noise}}&\multirow{2}{*}{\makecell[c]{Decentralized QKL /w $\tilde{p}_{node2}=0.5 \%$}} & \multirow{2}{*}{\makecell[c]{1250}} & \multirow{2}{*}{\makecell[c]{82.78\%\\(avg. w/o node2)}} & \multirow{2}{*}{\makecell[c]{100.00\%\\(avg. w/o node2)}} & \multirow{2}{*}{\makecell[c]{94.58\%\\(avg. w/ node2)}}\\ 
         & & & & & \\
        &Centralized QKL w/ $\tilde{p}=0.5 \%$ & 5650 & \rule[0pt]{1em}{0.4pt} & \rule[0pt]{1em}{0.4pt} & 98.33\%\\ 
        &Local QKL /w $\tilde{p}_{node2}=0.5 \%$ & 1250 & 71.43\% & 53.33\% & 73.33\%\\ 
    \midrule
        \multirow{2}{*}{\makecell[c]{Quite\\High Noise}}&\multirow{2}{*}{\makecell[c]{Decentralized QKL /w $\tilde{p}_{node2}=5 \%$}} & \multirow{2}{*}{\makecell[c]{1500}} & \multirow{2}{*}{\makecell[c]{100.00\%\\(avg. w/o node2)}} & \multirow{2}{*}{\makecell[c]{82.22\%\\(avg. w/o node2)}} & \multirow{2}{*}{\makecell[c]{95.00\%\\(avg. w/o node2)}}\\
        & & & & & \\
    \midrule 
        \multirow{2}{*}{\makecell[c]{Gaussian\\Attack}}&No Defense w/ $\tilde{p}_{all}=0.05 \%$& \rule[0pt]{1em}{0.4pt} & 59.47\% (avg.) & 49.15\% (avg.) & 51.60\% (avg.)\\
        &RDQKL w/ $\tilde{p}_{all}=0.05 \%$& 2050 & 100.00\% (avg.) & 92.78\% (avg.) & 97.78\% (avg.)\\
    \midrule
        \multirow{2}{*}{\makecell[c]{Sign-flippin\\Attack}}&No Defense w/ $\tilde{p}_{all}=0.05 \%$& \rule[0pt]{1em}{0.4pt} & 84.95\% (avg.) & 51.17\% (avg.) & 53.57\% (avg.)\\ 
        &RDQKL w/ $\tilde{p}_{all}=0.05 \%$& 1050 & 100.00\% (avg.) & 72.22\% (avg.) & 88.89\% (avg.)\\
    \bottomrule
    \end{tabular}
\end{table*}

\begin{table*}[ht]
    \centering
\caption{Experiment results on dataset 2 under noise environment}
\label{tab2}
    \begin{tabular}{cccc}
    \toprule
    \textbf{Scenario} &\textbf{Experiment Setting} & \textbf{Iteration Times} & \textbf{Score3}\\
    \midrule 
        \multirow{2}{*}{\makecell[c]{Normal Noise}}&Decentralized QKL w/ $\tilde{p}_{all}=0.05\%$ & 270  & 88.33\% (avg.)\\
        &Centralized QKL w/ $\tilde{p}=0.05 \%$ & 800 & 81.67\%\\
    \midrule
        \multirow{2}{*}{\makecell[c]{Relatively High Noise}}&Decentralized QKL w/ $\tilde{p}_{node2}=0.2 \%$ &  320 & 85.83\%  (avg.) \\ 
        &Centralized QKL w/ $\tilde{p}=0.2 \% $ & 1500 & 81.67\%\\
    \midrule
        \multirow{2}{*}{\makecell[c]{Quite High Noise}}&\multirow{2}{*}{\makecell[c]{Decentralized QKL /w $\tilde{p}_{node2}=5 \%$}} & \multirow{2}{*}{\makecell[c]{300}} & \multirow{2}{*}{\makecell[c]{85.00\% (avg. w/o node2)}}\\
        & & & \\
    \midrule 
        \multirow{2}{*}{\makecell[c]{Gaussian Attack}}&No Defense w/ $\tilde{p}_{all}=0.05 \%$ & \rule[0pt]{1em}{0.4pt}  & 48.44\% (avg.)\\
        &RDQKL w/ $\tilde{p}_{all}=0.05 \%$ & 270 & 85.00\% (avg.)\\
    \midrule
        \multirow{2}{*}{\makecell[c]{Sign-flippin Attack}}&No Defense w/ $\tilde{p}_{all}=0.05 \%$ & \rule[0pt]{1em}{0.4pt} & 82.50\% (avg.)\\ 
        & RDQKL w/ $\tilde{p}_{all}=0.05 \%$ & 290 & 83.89\% (avg.)\\ 
    \bottomrule
    \end{tabular}
\end{table*}

We will evaluate the performance of Decentralized QKL algorithm through experiments on both artificial dataset as shown in Fig. \ref{data fig1} and real-world dataset as shown in Fig. \ref{data fig2}. Experiments involve manual partitioning and random partitioning the dataset, revealing favorable performance of the algorithm in both cases. Two different communication topologies are also chosen to assess the overall performance. The first topology is a ring network, where each quantum computing unit communicates only with its two adjacent neighbors. The second topology is a fully connected network, where each node exchanges information with all other nodes.

When randomly partitioning the dataset, the primary goal is typically to accelerate training using the decentralized method. Therefore, for Dataset 2, we focus on evaluating the effectiveness of decentralized methods in speeding up the training process. In contrast, for the unevenly partitioned Dataset 1 obtained via manual division, our evaluation not only considers the acceleration of training through the decentralized method but also investigates the performance when using the decentralized method for privacy protection. As shown in Table~\ref{tab1} and~\ref{tab2}, four evaluation metrics are used: iteration times, Score1, Score2, and Score3. Specifically, Score1 denotes the accuracy achieved by model training on the local training set and testing on the local test set; Score2 represents the accuracy obtained by model training on the local training set but testing on the whole test set; Score3 refers to the accuracy when both model training and testing are conducted on the whole dataset. Iteration times measure the speed of convergence during training. Score1 reflects the model's performance on local data when trained solely on local information, Score2 captures the model's generalization ability to unseen data using only local training, and Score3 indicates the overall model performance when decentralized methods are employed to accelerate training. Additionally, experiments are also conducted with the centralized QKL based on all data for comparison.

The feature mapping circuit is a stack of $8$ layers of the circuit as explained in Fig. \ref{fig_c}. The number of the working qubits $n$ is $5$ and the embedding layer alternately embeds the binary features. That is, $x_1$ is embedded in the $1^{st}$, $3^{rd}$, and $5^{th}$ qubits, and $x_2$ is embedded in the $2^{nd}$ and $4^{th}$ qubits. The noise channel is depolarizing noise with parameter \(\tilde{p}\). The weighted average aggregation method is used in the algorithm. $\tau$ in $\text{Clip}(\theta_{(i,j)}^{(k+\frac{1}{2})}, \tau)$ is chosen as 0.5 and 0.05 for Gaussian attack and Sign-flipping attack respectively. The specific implementation of the experiments are based on Pennylane framework \cite{pennylane}. The backend using to simulate the quantum computer is the default.mixed of Pennylane and the method of differentiation to use is backprop.

\subsection{Decentralized Performance Under Noise}

\subsubsection*{Normal Noise}

As shown in Tab~\ref{tab1} and~\ref{tab2}, when the noise level remains within a normal range, the decentralized QKL achieves significantly faster training compared to the centralized QKL. This improvement is primarily due to the involvement of more quantum computing units. Moreover, the decentralized approach yields comparable or even superior model performance. When using for privacy preserving, as demonstrated in Tab~\ref{tab1}, the decentralized method leverages global information, compared to using only local data, the decentralized method is equivalent to using global information, so that the model can obtain better generalization performance on global data.

\subsubsection*{Relatively High Noise}

When the noise level of individual quantum computing units is relatively high, the discriminative power of the quantum kernel may degrade, as shown by the performance of Local QKL in Tab~\ref{tab1}. However, as illustrated in Tab~\ref{tab1} and~\ref{tab2}, the decentralized QKL continues to accelerate the overall training process without being affected by the high noise in certain units. By leveraging global information, even quantum computing units subject to higher noise levels can still achieve improved performance at the global level. When using for privacy preserving, as shown in Tab~\ref{tab1}, quantum units operating under normal noise conditions remain unaffected by those with higher noise, demonstrating the robustness of the decentralized approach in heterogeneous environments.

\subsubsection*{Quite High Noise}

When certain quantum computing units suffer from significantly high noise levels while others operate under normal noise conditions, the quantum kernels associated with the high-noise units contribute minimally to the training due to their small gradient magnitudes. As a result, they have little to no adverse impact on the training of other quantum kernels. As shown in Tab~\ref{tab1} and~\ref{tab2}, although the overall training speed decreases slightly compared to the case where all units have normal noise levels, it still substantially outperforms the centralized QKL. Furthermore, quantum computing units with normal noise levels consistently achieve strong model performance, both when using for accelerating training and privacy preserving.

\subsection{Robustness Under Adversarial Attacks}

The Gaussian attack and Sign-flipping attack can significantly affect general decentralized QKL. As shown in Tab~\ref{tab1} and~\ref{tab2}, the Gaussian attack greatly disrupts training performance across various scenarios. However, when robust defense strategies are employed, the impact of the Gaussian attack can be effectively mitigated. Although the Sign-flipping attack is generally less destructive than Gaussian attacks, their effect is still non-negligible. When using the \text{Clip} robust strategy, the Sign-flipping attack can be partially alleviated, though the mitigation is not as effective as in the case of the Gaussian attack.  It is also worth noting that robust decentralized training strategies may lead to longer training times. Nonetheless, they still offer significantly better performance compared to centralized methods.

\section{Conclusion}
\label{sec:conclusion}
In this paper, we analyze the impact of noise on QKL and study the robustness of decentralized QKL to the noise. By integrating robust decentralized optimization technique, we propose a RDQKL framework that jointly addresses noise degradation and adversarial threats in a unified manner. Empirical evaluations on synthetic and real-world datasets show that the proposed approach retains great efficiency and strong classification performance even under severe depolarizing noise and deliberate adversarial corruption, highlighting the potential of decentralized quantum kernel methods to operate securely and effectively on near-term quantum hardware. These results pave the way for broader deployments of QML in real-world and large-scale applications, where heterogeneous hardware conditions and untrusted participants are the norm, and suggest promising avenues for future research on scalable, noise-aware, and security-focused quantum algorithms. In future work, we will try to apply the proposed method to recent quantum hardware and solve practical problems in real-world scenarios such as power systems \cite{ma2023grid, 10380506, 10459229}.
% Our future works will focus on applying the proposed quantum algorithms to realistic power system applications \cite{10380506,10459229}.

\section*{Data Availability}
The code supporting the findings of this research is available on GitHub at the following repository: \href{https://github.com/Leisurivan/RDQKL}{https://github.com/Leisurivan/RDQKL}. This repository includes the scripts and data required to reproduce the results presented in this paper.

\section*{Acknowledgment}

The authors would like to thank Bing Liu (Zhejiang University) and Aaron Sander (Technical University of Munich) for useful discussions and suggestions. SY thanks UK Research and Innovation Guarantee Postdoctoral Fellowship (project: EP/Y029631/1).

% \clearpage

% \appendices
% \section{Proof of Proposition 1}\label{A}

% \clearpage

\balance

\bibliographystyle{ieeetr}
\bibliography{references}

\end{document}